\documentclass{article}

\usepackage{arxiv}

\usepackage[utf8]{inputenc} 
\usepackage[T1]{fontenc}    
\usepackage{hyperref}       
\usepackage{url}            
\usepackage{booktabs}       
\usepackage{amsfonts}       
\usepackage{nicefrac}       
\usepackage{microtype}      
\usepackage{lipsum}		
\usepackage{graphicx}
\usepackage{natbib}
\usepackage{doi}
\usepackage{multirow}  
\usepackage{graphicx}  
\usepackage{tabularx}
\usepackage{amsmath}

\title{Patient-Level Multimodal Question Answering from Multi-Site Auscultation Recordings} 

\author{{\hspace{1mm}Fan Wu \thanks{Corresponding author \texttt{fanwu@ethz.ch}}}\\
	Agentic Systems Lab, ETH Zurich\\
    Switzerland \\
	\And
	{\hspace{1mm}Tsai-Ning Wang} \\
	Eindhoven University of Technology \\
    Netherlands \\
	\And
	{\hspace{1mm}Nicolas Zumarraga} \\
	Agentic Systems Lab, ETH Zurich\\
    Switzerland \\
	\And
	{\hspace{1mm}Ning Wang} \\
	Agentic Systems Lab, ETH Zurich\\
    Switzerland \\
	\And
	{\hspace{1mm}Markus Kreft} \\
	Agentic Systems Lab, ETH Zurich\\
    Switzerland \\
	\And
	{\hspace{1mm}Kevin O'Sullivan} \\
	Agentic Systems Lab, ETH Zurich\\
    Switzerland \\
	\And
	{\hspace{1mm}Elgar Fleisch} \\
	Agentic Systems Lab, ETH Zurich\\
    Centre for Digital Health Interventions, ETH Zurich\\
    Centre for Digital Health Interventions, University of St. Gallen \\
    Switzerland \\
	\And
	{\hspace{1mm}Oliver Aalami} \\
	Stanford Mussallem Center for Biodesign, Stanford University\\
    USA \\
	\And
	{\hspace{1mm}Paul Schmiedmayer} \\
	Stanford Mussallem Center for Biodesign, Stanford University\\
    USA \\
	\And
	{\hspace{1mm}Robert Jakob \thanks{Equal contribution as senior authors} } \\
	Agentic Systems Lab, ETH Zurich\\
    Switzerland \\
	\And
	{\hspace{1mm}Patrick Langer \footnotemark[2]} \\
	Agentic Systems Lab, ETH Zurich\\
    Stanford Mussallem Center for Biodesign, Stanford University \\
    Centre for Digital Health Interventions, ETH Zurich \\
    Switzerland \\
}

\renewcommand{\shorttitle}{Patient-Level Multimodal Question Answering from Multi-Site Auscultation Recordings}

\begin{document}
\maketitle

\begin{abstract}
Auscultation is a vital diagnostic tool, yet its utility is often limited by subjective interpretation. While general-purpose Audio-Language Models (ALMs) excel in general domains, they struggle with the nuances of physiological signals. We propose a framework that aligns multi-site auscultation recordings directly with a frozen Large Language Model (LLM) embedding space via gated cross-attention. By leveraging the LLM’s latent world knowledge, our approach moves beyond isolated classification toward holistic, patient-level assessment. On the CaReSound benchmark, our model achieves a state-of-the-art 0.865 F1-macro and 0.952 BERTScore. We demonstrate that lightweight, domain-specific encoders rival large-scale ALMs and that multi-site aggregation provides spatial redundancy that mitigates temporal truncation. This alignment of medical acoustics with text foundations offers a scalable path for bridging signal processing and clinical assessment. \footnote{This manuscript has been submitted to Interspeech 2026 for peer review.}

\end{abstract}


\section{Introduction}

Auscultation remains a clinical cornerstone for monitoring heart and lung function~\cite{cook2022body}, yet its diagnostic utility is limited by subjective interpretation and inter-observer variability~\cite{AvilesSolis2017}. While machine learning (ML) methods enable automated detection of pathologies from auscultation signals, most approaches frame the task as classification, reducing complex physiological signals to discrete labels~\cite{springer2016logistic, rocha2019respiratory}. Such formulations offer limited flexibility for clinical use. Question answering (QA) provides an alternative paradigm by allowing models to respond to natural language queries about medical data~\cite{jin2021disease}. In auscultation analysis, QA allows clinicians to ask targeted questions about patient recordings.

LLMs have demonstrated strong capabilities in QA and generating domain-expert clinical rationales~\cite{brown2020language}. Their success has inspired extending generative models beyond text to other modalities, including ALMs that integrate acoustic signals with LLMs, e.g., via cross attention to provide audio-conditioned textual interpretations of clinical sounds~\cite{kong2024audio}.

However, applying ALMs to auscultation QA presents several challenges. First, although general ALMs support audio inputs, they are primarily trained on speech or environmental sounds. In contrast, medical auscultation contains domain-specific signals with subtle pathological patterns (e.g., murmurs or crackles) that are often obscured by noise or overlapping physiological signals~\cite{ren2024comprehensive, moberg2025lung}. Second, while medical auscultation signals are temporally structured and relatively long (e.g., up to 30~s), yet current approaches rely on short-window segmentation (2–5~s)~\cite{li2026advances}. This fragmentation may lose the ``rhythmic context'' of full respiratory or cardiac cycles. Third, auscultation requires integrating evidence across multiple anatomical sites. Although datasets such as PhysioNet provide multi-site recordings, many methods still operate on single recordings and ignore cross-site relationships~\cite{reyna2022heart}.

To address these challenges, we propose a patient-level multimodal clinical QA framework that bridges physiological audio signals and multimodal generative modeling for holistic auscultation analysis. Inspired by Flamingo’s gated cross-attention~\cite{alayrac2022flamingo}, our architecture leverages the LLM’s latent world knowledge by integrating multi-site recordings (up to 30~s) as a native modality to produce grounded free-text answers. On the CaReSound benchmark, our model establishes a new state-of-the-art, achieving a 42.6\% Contains-Match (+7.75 points) and 0.865 F1-macro (+1.9 points) over the CaReAQA baseline, with a 0.952 BERTScore. Beyond raw performance, we demonstrate that lightweight, domain-specific encoders using raw-waveform tokenization can rival large-scale pretrained ALMs. We provide a systematic study of temporal and spatial dynamics, showing that multi-instance aggregation provides spatial redundancy that enhances robustness to signal truncation. Our framework explicitly supports physiological audio, providing novel insights into grounding language models in the medical acoustic domain. Code will be released upon acceptance.

\section{Related Work}

\subsection{From Acoustic Classification to Medical QA}

Traditional analysis of medical acoustics primarily relies on supervised classification for heart and lung sounds~\cite{partovi2024review}. While effective for predefined patterns, these models are task-specific and lack the flexibility for natural language interaction~\cite{sfayyih2023review}. Although general ALMs like Audio Flamingo~\cite{kong2024audio} and Qwen-Audio~\cite{chu2023qwen} enable audio-text interaction, they are optimized for speech or environmental sounds rather than the subtle patterns of physiological signals.

In the medical domain, while medical visual QA is well-established for imaging~\cite{van2023open}, physiological audio remains underrepresented. Domain-specific efforts like RespLLM focus on respiratory classification~\cite{zhang2024respllmunifyingaudiotext}, while OPERA learns robust respiratory encoders, it lacks a generative framework for open-ended clinical QA~\cite{zhang2024towards}. The CaReAQA represents an initial step toward audio-grounded open-ended QA ~\cite{wang2025careaqa}; however, it relies on segmented, single-instance recordings, which may disrupt global temporal patterns and fail to capture the multi-site nature of clinical auscultation. Our work extends this by aligning multi-site physiological signals with the latent world knowledge of frozen LLMs, enabling holistic, patient-level clinical assessment through grounded QA.

\subsection{Audio Representation}

To process continuous auscultation signals with Transformer-based models, audio need be converted into discrete representations, typically via temporal patching~\cite{liu2023visual} or embeddings from self-supervised models like Wav2Vec 2.0~\cite{baevski2020wav2vec2}. However, these encoders often become computationally prohibitive when processing the long-duration recordings required to preserve physiological context. While hierarchical architectures like MEGABYTE utilize chunking to mitigate this~\cite{yu2023megabyte}, they are primarily optimized for linguistic data and often fail to exploit the quasi-periodic, non-linguistic structures inherent in physiological signals.

Furthermore, clinical auscultation is inherently a multi-instance process, requiring clinicians to integrate acoustic evidence across multiple anatomical sites (e.g., aortic, pulmonic) rather than interpreting isolated clips. This workflow aligns with MIL, where a patient is treated as a ``bag'' of recordings informed by localized ``instances'' of pathology~\cite{duggento2021classification}.

To bridge the need for long-context modeling with multi-site fusion, we employ latent compression via Perceiver Resamplers~\cite{alayrac2022flamingo}. By projecting long, multi-instance sequences into a fixed set of latent vectors, we mitigate the token bottleneck while preserving both local fine-grained features and global temporal patterns. This allows our framework to scale to comprehensive patient-level inputs without the memory constraints typical of standard Transformer architectures.

\section{Methods}


\subsection{Dataset}
\label{sec:dataset}

We use \textbf{CaReSound}, a benchmark for medical QA on cardiac and respiratory auscultation audio~\cite{wang2025careaqa}, built from several public datasets including ICBHI~\cite{ICBHI17-Resp}, KAUH~\cite{KAUH21-Resp}, CirCor~\cite{CirCor22-Heart}, SPRSound~\cite{SPRSound22}, and ZCHSound~\cite{ZCHSound24-Heart}. Textual metadata and annotations are transformed into open-ended diagnostic dialogues using GPT-4o, simulating clinical scenarios where physicians analyze patient recordings and provide diagnostic feedback. Audio recordings and QA pairs are linked by patient ID, enabling \textit{patient-level assessment}. Datasets are categorized by structure: \textit{Single-instance (SIL)} datasets (KAUH, ZCHSound) contain one recording per patient, whereas \textit{MIL} datasets (ICBHI, CirCor, SPRSound) include recordings from multiple anatomical locations. Location metadata from file names is used as contextual input (e.g., CirCor provides aortic, pulmonary, tricuspid, and mitral valve recordings per patient). Thus, patient-level prediction requires aggregating information across recordings. Table~\ref{tab:dataset} summarizes the SIL and MIL datasets.

\begin{table}[h]
\centering
\caption{Auscultation dataset statistics}
\label{tab:dataset}
\begin{tabular}{lccccc}
Dataset & Patients & Audios & Audios/pt & QAs & QA/pt \\
\hline
CirCor & 942 & 3,163 & 3.4 & 3,284 & 3.5 \\
ICBHI & 126 & 920 & 7.3 & 20,728 & 164 \\
SPRSound & 292 & 2,683 & 9.2 & 5,029 & 17.2  \\
KAUH & 336 & 336 & 1.0 & 1,009 & 3.0 \\
ZCHSound & 1,259 & 1,259 & 1.0 & 2,527 & 2.0 \\
\hline
Total & 2,951 & 8,361 & - & 32,577 & - \\
\end{tabular}
\end{table}

\subsection{Audio Pre-processing}

The source datasets contain recordings with heterogeneous sampling rates (4--44.1~kHz). To ensure consistency, all audio is resampled to 16 kHz and converted to mono via channel averaging. Audio clips are truncated to a maximum duration of 30~s (480{,}000 samples), with shorter signals zero-padded to this length. Waveforms are normalized to zero mean and unit variance. Finally, to ensure compatibility with the encoder's patch-based architecture, signals are padded to the nearest multiple of the patch size (640 samples, 40~ms). While zero-padding introduces non-informative regions, the subsequent Perceiver Resampler compresses these representations into fixed latent tokens, mitigating the impact of the padding.

\subsection{Multimodal Architecture}
\label{sec:architecture}

\begin{figure*}[t]
  \centering
  \includegraphics[width=\textwidth]{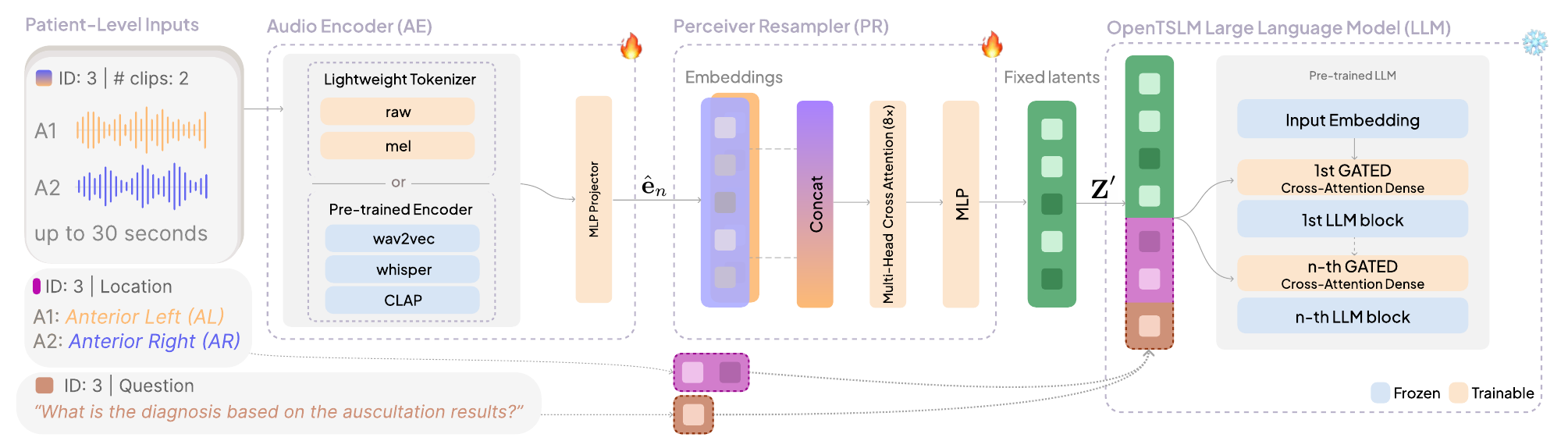}
  \caption{Architecture of AudioTSLM for patient-level multimodal clinical QA}
  \label{fig:pipeline}
\end{figure*}

Based on OpenTSLM-Flamingo architecture~\cite{planger2025opentslm}, we convert auscultation recordings into temporal token sequences, aggregate them into a patient-level latent representation, and inject this representation into a frozen LLM via gated cross-attention, as shown in Figure~\ref{fig:pipeline}.

\subsubsection{Audio Representation \& Encoder}

Each recording $\mathbf{x} \in \mathbb{R}^{L}$ is converted into a sequence of acoustic tokens $\mathbf{e}_n$ via one of two alternative tokenizers. The \textit{RawAudioTokenizer} segments the waveform into non-overlapping 40~ms patches, projected via 1-D convolution:
\begin{equation}
\mathbf{e}_n = \text{Conv1D}\big(\mathbf{x}_{nP:(n+1)P}; W_{\text{patch}}\big) + \mathbf{p}_n
\end{equation}
where $W_{\text{patch}}$ is the kernel and $\mathbf{p}_n$ is a learnable positional embedding. This transformation allows the model to process auscultation signals as sequential tokens, analogous to words in a sentence. Alternatively, the \textit{MelSpectrogramTokenizer} processes a log-mel spectrogram $\tilde{\mathbf{S}}$ through a 2-D CNN with frequency-wise pooling:
\begin{equation}
\mathbf{e}_n = \text{AvgPool}_{\text{freq}}\big(\text{CNN}_{2\text{D}}(\tilde{\mathbf{S}})\big)_n + \mathbf{p}_n
\end{equation}

Optionally, pretrained encoders (i.e., Wav2Vec2, Whisper~\cite{radford2023robust}, CLAP~\cite{elizalde2023clap}) can replace the tokenizers. Regardless of the encoder choice, the resulting token sequences are projected into a shared $D_{\text{proj}}$-dimensional space via a LayerNorm-GELU Multi-Layer Perceptron (MLP) to ensure a uniform latent representation for downstream conditioning:

\begin{equation}
\hat{\mathbf{e}}_n = \text{GELU}\Big(\text{LayerNorm}(\mathbf{e}_n) W_{\text{proj}} + \mathbf{b}_{\text{proj}}\Big)
\end{equation}

\subsubsection{Perceiver Resampler \& Multi-Instance Learning}

For a patient with $M$ recordings, tokens are zero-padded to a uniform length $N_{\max}$ and flattened into a multi-instance matrix $\mathbf{X} \in \mathbb{R}^{(M \cdot N_{\max}) \times D}$. To obtain a fixed-length patient-level representation, we employ a \textit{Perceiver Resampler} where $K$ learnable latent queries $\mathbf{Z} \in \mathbb{R}^{K \times D}$ attend jointly across all temporal and clip dimensions:
\begin{equation}
\mathbf{Z}' = \text{softmax} \left( \frac{\mathbf{Z} W_Q (\mathbf{X} W_K)^\top}{\sqrt{D_k}}\right) \mathbf{X} W_V
\label{eq:perceiver}
\end{equation}
Instead of averaging or concatenating recordings, this cross-attention mechanism acts as an adaptive pooling operator, enabling the model to capture \textit{cross-site interactions} (e.g., AV--MV co-variation) while producing a fixed-size latent representation $\mathbf{Z}'$  regardless of the encoder choice.

\subsubsection{Gated Cross-Attention \& Multi-Modal Fusion}

For each recording, site metadata (e.g., ``AV'') is embedded as natural language and paired with a structural \textless audio\textgreater\ token. The patient-level prompt follows the format: \texttt{[Instruction] \dots \textless audio\textgreater\ [Site] \dots\ Question: [Q] Answer: [A]}. Multimodal integration occurs via gated cross-attention layers interleaved between frozen Transformer blocks. At layer $\ell$, the LLM representations $\mathbf{H}^{(\ell)}$ act as queries to the Perceiver latents $\mathbf{Z}'$, which serve as keys and values:
\begin{equation}
\mathbf{H}^{(\ell)} \leftarrow \mathbf{H}^{(\ell)} + \tanh(\alpha_{\ell}) \cdot \text{CrossAttn}(\mathbf{H}^{(\ell)}, \mathbf{Z}')
\label{eq:gatedcross}
\end{equation}
The learnable $\tanh(\alpha_{\ell})$ gating parameter regulates the influence of acoustic features. It preserves pretrained language priors while progressively aligning the medical acoustic patterns with textual embeddings.

\section{Experiments}
\label{sec:experiments}

Our experiments are designed to identify the optimal configuration for grounding a LLM in multi-site physiological audio. We first establish a baseline for patient-level QA, then systematically evaluate the two primary factors governing clinical assessment: the acoustic representation (how the model encodes audio) and temporal context (the duration of the input signal).

\subsection{Patient-Level Question Answering}

Our primary study focuses on open-ended and binary clinical QA and we evaluate our framework on the CaReSound dataset. We use cross-dataset stratified splits to prevent data leakage, resulting in 2,064 / 440 / 447 patients (22,882 / 3,199 / 6,496 QA pairs) for training, validation and testing. Our framework utilizes the \textit{RawAudioTokenizer} integrated with a frozen Meta-LLaMA-3.2-1B backbone via trainable gated cross-attention adapters, with approximately 1.4B trainable parameters. Experiments were conducted on NVIDIA RTX PRO 6000 Blackwell GPUs (97,887~MiB VRAM). Models are trained with AdamW optimizer with an effective batch size of 16 and component-specific learning rates ($5 \times 10^{-6}$ for encoders and $1.5 \times 10^{-5}$ for adapters). To mitigate dataset imbalance, we employ balanced sampling. At the end, we evaluate performance on the held-out test set. Open-ended QA is measured using Contains-Match Accuracy, ROUGE-L F1, METEOR, and BERTScore F1 to capture both lexical and semantic similarity, while binary questions are assessed via Accuracy, F1, Sensitivity, and Specificity. More details are provided in the supplementary material.

\subsection{Audio Representation \& Audio Encoders}

To determine the most effective bridge between raw auscultation and LLM embeddings, we compare five encoder variants: two lightweight, scratch-trained front-ends (RawAudio, Mel-Spectrogram) and three large-scale pretrained models (Wav2Vec2, Whisper, CLAP). This comparison quantifies whether specialized physiological tokenization can rival or outperform generic, high-capacity audio encoders for clinical tasks. Training a full model requires approximately 4~h for lightweight tokenizers and 8~h for pretrained encoders, while inference requires around 8~h.

\subsection{Temporal Context Modeling}

Auscultation requires capturing transient pathological sounds, such as murmurs or crackles, which may be missed in short clips. We assess the impact of temporal context by reducing the maximum audio duration (from 30~s to 20~s and 10~s) using the RawAudioTokenizer. This study identifies the point of diminishing returns for sequence length and determines the minimum duration required for reliable patient-level inference.

\section{Results}

\subsection{Patient-Level Question Answering}

Table~\ref{tab:comparision} (top) compares our framework against zero-shot foundational ALMs (Audio-Flamingo3, Qwen2-Audio, and Qwen2.5-Omni) and fine-tuned baseline (CaReAQA). Our 1.4B model consistently outperforms foundational ALMs pretrained on large amounts of audio data, highlighting the need for domain-specific fine-tuning. Notably, our framework improves upon fine-tuned CaReAQA across all metrics, particularly in generative grounding: we achieve a Contains-Match Accuracy of 42.6\% (+7.75 points; 22\%), ROUGE-L of 0.673 (+0.061; 10\%), and METEOR of 0.643 (+0.133; 26\%), and BERTScore of 0.952. While binary classification improvements are more moderate, increasing F1-macro from 0.846 to 0.865 (+1.9 points). The substantial gains in open-ended metrics indicate significantly better textual alignment and answer grounding. 

\begin{table*}
\centering
\footnotesize
\caption{Performance comparison across state-of-the-art ALMs, different audio encoders and different temporal context lengths}
\label{tab:comparision}
\begin{tabular}{llcccccc}
\toprule
& Approach & Yes/No Acc (\%) & Yes/No F1 & Contains-Match (\%) & ROUGE-L & METEOR & BERTScore \\
\midrule
\multirow{5}{*}{\rotatebox{90}{Model}}
& Audio-Flamingo3 (8B) & 85.00 & 0.545 & 5.51 & 0.2883 & 0.2842 & 0.8851 \\
& Qwen2-Audio (7B) & 17.83 & 0.171 & 4.33 & 0.2951 & 0.3592 & 0.8996 \\
& Qwen2.5-Omni (0.5B) & 84.62 & 0.523 & 6.74 & 0.1351 & 0.2276 & 0.8648 \\
& CaReAQA (3B) & 93.12 & 0.846 & 34.85 & 0.6117 & 0.5107 & 0.9423 \\
& Ours (1.4B) & \textbf{93.50} & \textbf{0.865} & \textbf{42.60} & \textbf{0.6732} & \textbf{0.6432} & \textbf{0.9519} \\
\midrule
\multirow{5}{*}{\rotatebox{90}{Encoder}}
& Raw       & \textbf{93.50} & \textbf{0.8648} & 42.60 & \textbf{0.6732} & 0.6432 & \textbf{0.9519} \\
& Mel       & \textbf{93.50} & 0.8632 & 42.69 & 0.6687 & 0.6359 & 0.9507 \\
& CLAP      & 93.44 & 0.8630 & \textbf{43.13} & 0.6726 & \textbf{0.6453} & 0.9513 \\
& Whisper   & \textbf{93.50} & 0.8646 & 42.47 & 0.6673 & 0.6396 & 0.9509 \\
& Wav2Vec   & 93.21 & 0.8564 & 42.38 & 0.6643 & 0.6349 & 0.9503 \\
\midrule
\multirow{3}{*}{\rotatebox{90}{Context}}
& 30s & \textbf{93.5} & \textbf{0.865} & \textbf{42.6} & \textbf{0.673} & \textbf{0.643} & \textbf{0.952} \\
& 20s & 92.0 & 0.826 & 41.3 & 0.655 & 0.623 & 0.948 \\
& 10s & 91.1 & 0.804 & 39.5 & 0.639 & 0.607 & 0.946 \\
\bottomrule
\end{tabular}
\end{table*}

\subsection{Audio Encoder Evaluation}

As shown in Table~\ref{tab:comparision} (middle), the model demonstrates high robustness to the choice of audio representation, though subtle performance trade-offs emerge. The \textit{RawAudioTokenizer} (raw waveform) achieves the highest F1-macro (0.865), ROUGE-L (0.673), and BERTScore (0.952), validating that lightweight, scratch-trained front-ends can provide superior semantic alignment for physiological audio. While CLAP embeddings yield marginal gains in Contains-Match (43.1\%) and METEOR (0.645), suggests slightly better content coverage, large-scale encoders like Wav2Vec2 slightly underperform across both binary and generative tasks. These findings confirm that raw waveform tokens offer an efficient and effective representation for clinical QA, rivaling high-capacity, general-purpose audio encoders.

\subsection{Temporal Context Evaluation}

As shown in Table~\ref{tab:comparision} (bottom), performance consistently declines as context is reduced. The 30~s configuration achieves the highest overall scores, including a 42.6\% Contains-Match and 0.865 F1-macro. In contrast, 10~s segments incur substantial reductions, particularly in binary metrics (up to 10\% decline). 

A granular analysis (Supplementary Table) reveals that the necessity of temporal context is dataset-dependent. In MIL datasets (CirCor, ICBHI, SPRSound), the model leverages patient-level aggregation across multiple sites to partially compensate for shorter clips; for example, SPRSound accuracy remains above 90\% despite truncation. However, for SIL datasets (KAUH, ZCHSound), full-length audio is critical. In KAUH, ROUGE-L drops sharply from 0.845 to 0.542 when truncated to 10~s, and ZCHSound F1-macro falls from 0.980 to 0.864. These results confirm that while multi-site aggregation provides robustness, full temporal context is essential to maintain high semantic and classification accuracy in single-recording scenarios.

\section{Discussion}

Our framework demonstrates a strong ability to translate acoustic signals into grounded textual explanations. Qualitative analysis reveals high BERTScore values even with concise predictions, indicating that the model captures core semantic content without strictly mimicking expert-annotated lengths. Our results further show that lightweight, domain-specific encoders, specifically those utilizing raw waveform tokenization, suffice for high-quality clinical QA, suggesting that large-scale pretrained encoders are not strictly necessary for medical auscultation. A key insight is the interplay between temporal context and aggregation: in multi-site datasets, the model leverages spatial redundancy across anatomical locations to maintain robustness against temporal truncation. 

However, the ``black-box'' nature of deep learning requires rigorous validation to avoid the false hope of ungrounded explainability. Our model addresses this by aligning free-text answers with signal-level evidence, though future work must focus on mitigating dataset bias and ensuring interpretability across diverse clinical settings. Ultimately, these findings suggest that audio-grounded language models can bridge the gap between signal processing and clinical interpretation, offering promising applications for telemedicine screening and diagnostic standardization in resource-limited environments.

\bibliographystyle{unsrtnat}
\bibliography{references}  






\end{document}